\documentclass[12pt]{article}
\usepackage{a4wide}
\usepackage{graphics}
\usepackage{graphicx}
\usepackage{amsmath}
\usepackage{booktabs}
\usepackage{float}
\usepackage{color}
\usepackage{cite}

\newcommand{\Jp}{J/\psi}

\newcommand{\GeV}{\mbox{\,GeV}}

\newcommand{\mub}{\mbox{\,$\mu$b}}

\newcommand{\pb}{\mbox{\,pb}}

\newcommand{\mrad}{\mbox{\,mrad}}

\begin{document}

\thispagestyle{empty} 

\begin{center} 
{\bf\large Phenomenological study for the search of evidence\\
 for intrinsic charm at the COMPASS experiment}\\[1.0cm]
{\large Andrei~Gridin$^1$,  Stefan~Groote$^2$,
Alexey~Guskov$^1$ and Sergey~Koshkarev$^2$}\\[0.3cm]
$^1$ Joint Institute for Nuclear Research, 141980 Dubna, Russia\\[3pt]
$^2$ Institute of Physics, University of Tartu, 51010 Tartu, Estonia\\[3pt]
\end{center}

\vspace{0.2cm}
\begin{abstract}
In this paper we conduct a phenomenological study for the search of evidence
for the intrinsic charm mechanism in double $\Jp$ production at the COMPASS
experiment  using the CERN $\pi^-$ beam at $190\GeV/c$.  We also re-review the 
double $\Jp$ production data provided by the NA3 experiment using the 
CERN $\pi^-$ beam at $150$ and $280\GeV/c$ with incident on a platinum target.
\end{abstract}

\section{Introduction}
Almost four decades have passed since the intrinsic charm mechanism was
proposed~\cite{Brodsky:1980pb}, stating that heavy quarks are present in the
proton's wavefunction from the outset.

The existence of heavy quarks in the proton's light-front (LF) wavefunction at
a large LF momentum fraction $x$ is in fact predicted by QCD if one analyzes
the higher Fock states $|uud c\bar c\rangle$ and $|uud c\bar cc\bar c\rangle$
in the hadronic eigenstate, i.e.\ Fock states where the heavy quark pairs are
multi-connected to the valence quarks. LF wavefunctions, the eigensolutions of
the QCD LF Hamiltonian, are defined at fixed LF time $\tau=t+z/c$ and are thus
off-shell in the invariant mass. For example in QED, positronium has an
analogous Fock state $|e^+e^-\mu^+\mu^-\rangle$ due to the insertion of
light-by-light scattering in the positronium self-energy amplitude.
In such an ``intrinsic charm'' Fock state $|uudc\bar c\rangle$, the maximum
kinematic configuration occurs at minimum invariant mass where all quarks are
at rest in the hadron's rest frame, i.e., at equal rapidity in the moving
hadron. Equal rapidity implies $x_i\propto(m^2+{\vec k_\perp}^2)^{1/2}$ for
each quark, so that the heavy quarks in the Fock state carry most of the
hadron's LF momentum. The operator product expansion predicts that the
probability of intrinsic heavy-quark Fock states $|uud Q\bar Q\rangle$ 
scales as $1/m_Q^2$ due to the non-Abelian couplings of
QCD~\cite{Brodsky:1984nx,Franz:2000ee}.

Even though there is no clear observation of the mechanism, the baryonic
states $\Lambda_c(udc)$ and $\Lambda_b(udb)$ were both discovered at the
Intersecting Storage Rings at high values of the Feynman momentum fraction
$x_F$~\cite{Lockman:1979aj,Chauvat:1987kb,Bari:1991ty}.\footnote{In this paper 
$x_F$ denotes the Feynman-$x$ in the laboratory frame while
$x_F^*$ denotes the Feynman-$x$ in the center-of-mass system.}
The SELEX experiment provided the observation of a double charm baryon
$|ccd\rangle$ at a large mean value for $x_F$ and a relatively small mean
transverse momentum~\cite{Mattson:2002vu,Ocherashvili:2004hi}. In addition,
the NA3 experiment measured both the single-quarkonium hadroproduction
$\pi A\to \Jp X$~\cite{Badier:1983dg} and the double-quarkonium
hadroproduction $\pi A\to\Jp\Jp X$~\cite{Badier:1982ae} at high $x_F$. In
fact, all of the $\pi A\to \Jp \Jp X$ events were observed with a total value
of $x_F>0.4$.

As we show below, the NA3 kinematic features and the production rate can be a
result of a misunderstanding of the detector acceptance. Fortunately, the NA3
measurement can be confirmed or disproved at the COMPASS experiment using the
CERN $\pi^-$ beam at $190\GeV/c$.

In this paper we discuss kinematic features and analysis strategies for the
search of evidence for the intrinsic charm mechanism at the COMPASS experiment. 
We also estimate upper limits for the double $\Jp$ production cross section 
for perturbative QCD (pQCD) and for the intrinsic charm (IC) mechanism. 

\section{\label{acceptance}Revisiting the double $\Jp$ production at NA3}
Using the CERN pion beam at $150$ and $280\GeV/c$ to produce charm particles
with incident on hydrogen and platinum targets, the NA3 experiment provided
data on the production of the double $\Jp$ on platinum target in the kinematic
region $x_F^*(\Jp)>0$ with the respective production cross sections of
$18 \pm 8\pb$ and $30 \pm 10\pb$ per nucleon and the ratio
$\sigma(\Jp\Jp)/\sigma(\Jp)=(3\pm 1)\times 10^{-4}$ at both energies. 

\subsection{Acceptance of the NA3 detector}

In order to understand the NA3 data, we give a short overview over the layout
of the NA3 detector (see Ref.~\cite{Badier:1980kb} for a more complete
description). The NA3 detector consisted of a spectrometer with fixed targets
of liquid hydrogen (proton target, 30 cm long) and platinum (nuclear target,
6 cm long). The targets were located at a distance of 45 cm.

For the measurements the NA3 experiment used the beams of $p$, $\bar p$,
$K^\pm$, $\pi^\pm$ with intensities of $(3-5)\cdot 10^7$ particles per second.
To reduce the particle flux through the spectrometer, a beam dump absorbing
about 80\% of the charged particle flux was installed behind the platinum
target. The dump was made of a 1.5 m block of stainless steel and had a
conical core made of tungsten and uranium. The aperture angle of the cone
could be chosen as either 20 or 30 mrad. The stainless steel blocks surrounded
the conical core of the dump. Along the beam behind the dump, other parts of
the spectrometer were located such as a spectrometer magnet, tracking
detectors, counter hodoscopes and trigger hodoscopes.
At the end of the spectrometer an additional 1.8 m long iron absorber was
placed which played the role of a muon filter and reduced the low energy
particle background. Together with the other trigger hodoscopes, the trigger
hodoscope placed behind the muon filter had the purpose to select muons
originated from the targets. The trigger system imposed a condition on the
vertical component of the transverse momentum of the muons. To be registered,
a single muon had to satisfy the condition $p_T > 1\GeV/c$, while for
two muons in the event one had to have $p_T > 0.6\GeV/c$ for each muon.
Such requirements  eliminated a large fraction of pion and kaon decays and
rejected low mass resonances like  $\rho$, $\phi$ and $\omega$ mesons.

In order to be registered, muons had to pass more than 3 m of iron. As charged
particles, on this way they interacted with nucleons of the matter and spent
some of their energy for ionization and radiative effects. For example, by
passing through 3 m of iron a muon with energy of $150\GeV/c$ looses more than
$7.5\GeV$ of its energy. This leads to an acceptance notion which mostly
depends on the geometry of the setup, but also on the kinematics of the
particles.

\begin{center}\begin{figure}[t]
\centerline{\includegraphics[scale=0.55]{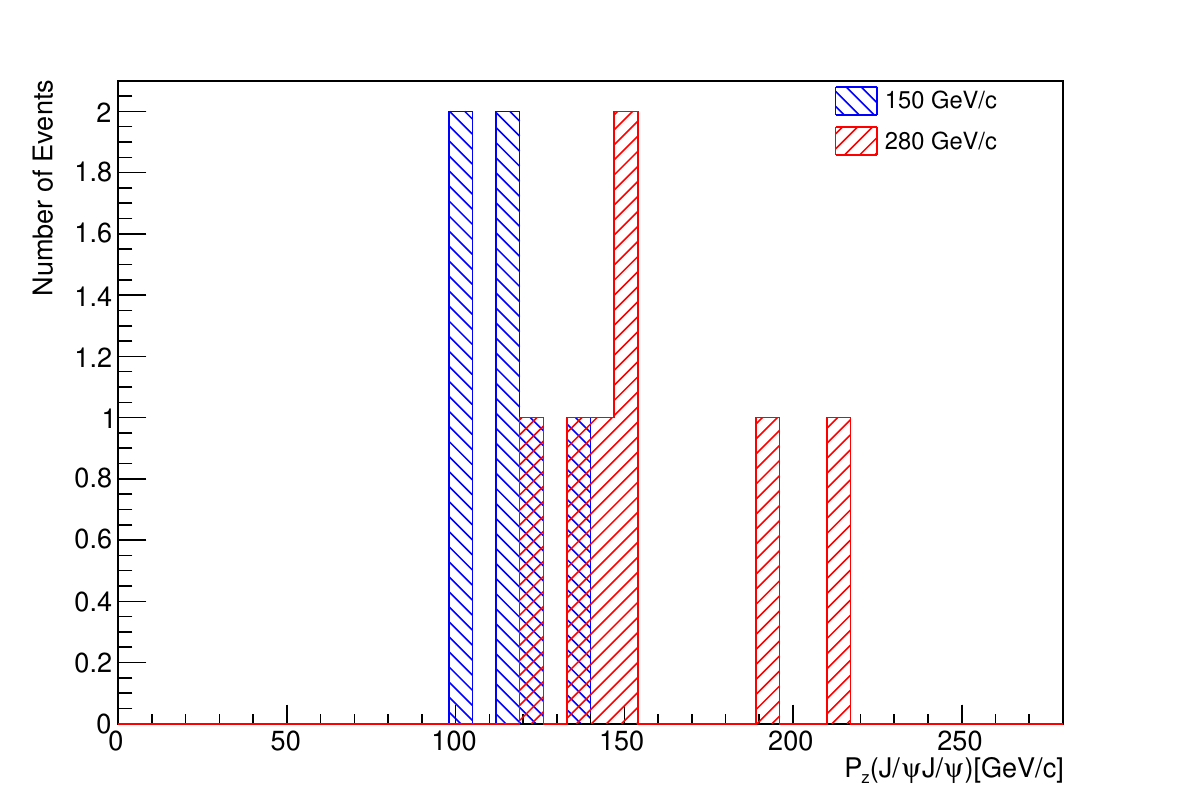}}
\caption{\label{fig:mom_na3}Momentum distributions of $\Jp$ pairs measured by
the NA3 collaboration.}
\end{figure}\end{center}

In the data analysis for single $\Jp$ selection a criterium $x_F^*>0$ was
used for both the $150\GeV$ and $280\GeV$ data samples. For $x_F^*<0$ the NA3
acceptance was dropping fast. This means that each of the $\Jp$ should
have had a minimal longitudinal momentum to pass the setup and to be detected.
For the $150\GeV$ beam this threshold was about $27\GeV/c$, and $39\GeV/c$ for
the $280\GeV$ beam. For the double $\Jp$ state these thresholds should be
multiplied roughly by two. Because the acceptance was dropping down near the
threshold, there was a low probability to detect an event with a momentum
close to the threshold. 
This means that it is not possible to detect a double $\Jp$ state with
$x_F \sim 27/150$ for $150\GeV$ and with $x_F \sim 39/280$ for $280\GeV$ 
{\it i.e.} $x_F < 0.4$ and $x_F < 0.3$ respectively, since low energy
muons will either be absorbed by the matter of the setup or rejected by the
trigger. In addition, because of the dropping of the acceptance, events
detected {\it de facto\/} by NA3 have values of $x_F$ larger than the
thresholds for both data samples (cf. Fig.~\ref{fig:mom_na3}). The estimate
for the NA3 setup acceptance for the double $\Jp$ production is done with a
Monte Carlo approach using pairs of uncorrelated $\Jp$'s. It is definitely
interesting to investigate the correctness of the acceptance obtained by such
a Monte Carlo simulation. 

\begin{figure}[t]
\includegraphics[scale=0.43]{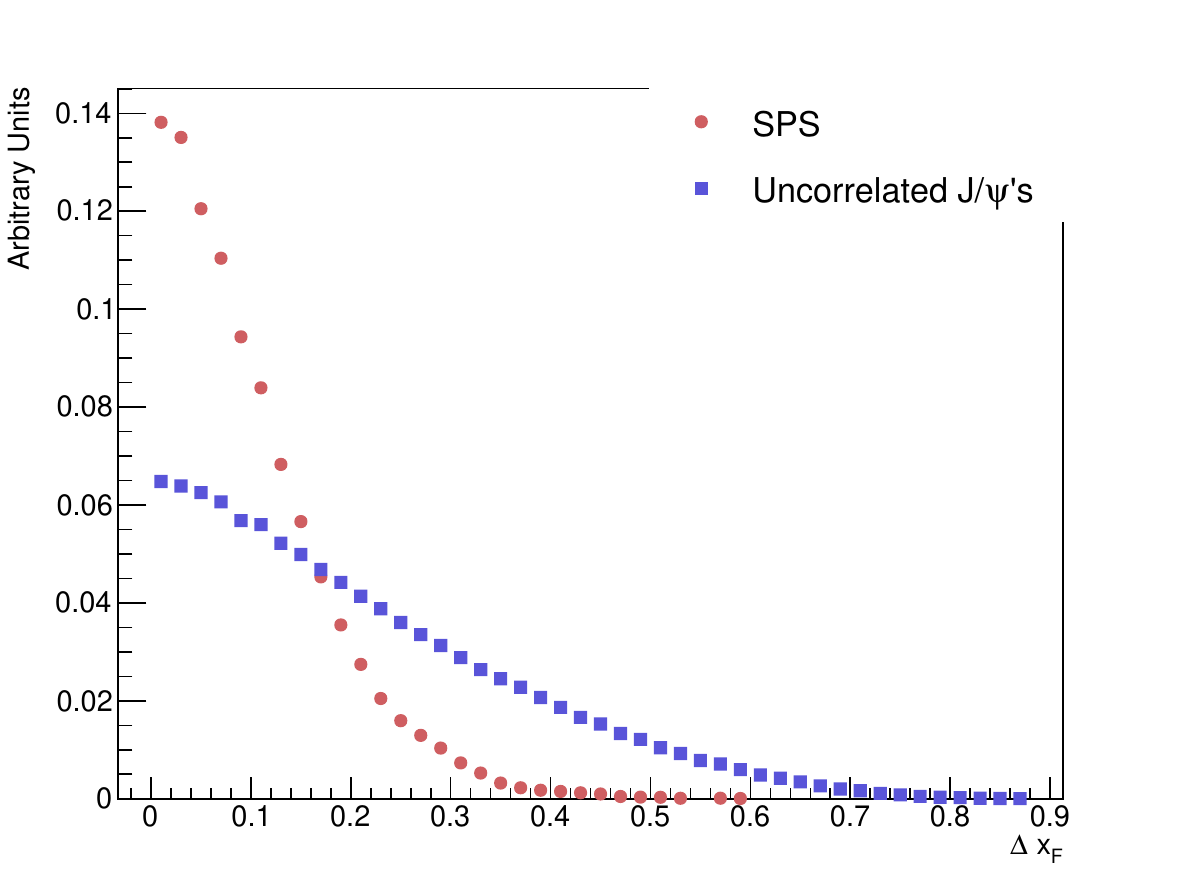}\kern-16pt
\includegraphics[scale=0.43]{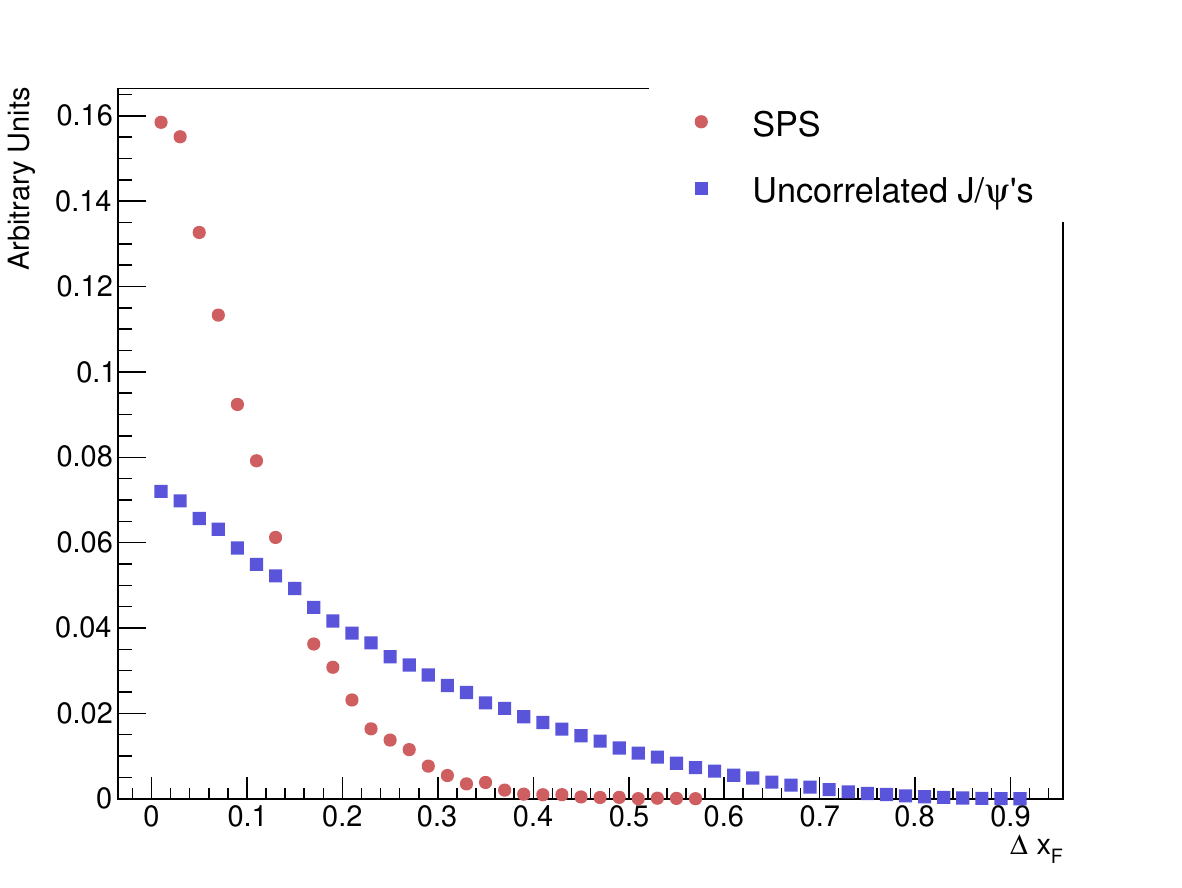}\kern-18pt
\caption{\label{fig:na3DPS} $\Delta x_F= |x_1(\Jp)-x_2(\Jp)|$ distributions
for the uncorrelated $\Jp$'s and SPS production mechanisms at the $150\GeV/c$
(left panel) and at the $280\GeV/c$ $\pi$ beam (right panel) at NA3.
$x_1(\Jp)$ and $x_2(\Jp)$ denote the $x_F$ for the first and the second $\Jp$,
respectively. The uncorrelated $\Jp$'s distribution is obtained using
Pythia~8~\cite{Sjostrand:2007gs}, and the SPS distribution is obtained using
HELAC-Onia~\cite{Shao:2012iz, Shao:2015vga}. All distributions are normalized
to unity.}
\end{figure}

By investigating the kinematic distributions (cf.\ Fig.~\ref{fig:na3DPS}), one
obtains a small difference in the momentum distribution between the $\Jp$ for
the Single Parton Scattering (SPS) mechanism and a higher momentum gap for
the uncorrelated $\Jp$'s sample. Such a gap in momentum and as the result of
this also in the Feynman-$x$ distributions could lead to the erroneous
interpretation of the NA3 acceptance. Indeed, keeping in mind that the $\Jp$
pair has to carry a minimum $x_F$ to be detected, a situation is possible
where one of the $\Jp$ does not cary enough momentum to be triggered. Due to
the larger momentum gap, the possibility of rejection can differ between SPS
and the uncorrelated sample. 
Concluding the above discussion we can say that the cross section values 
provided by the NA3 might be not fully correct and the high Feynman-$x$ region
can be interpreted solely by the detector acceptance.

\subsection{$\sigma(\Jp\Jp)$ production via Single Parton Scattering}

Utilizing the fact that the double $\Jp$ production cross section can be
normalized with a special choice of the composition $(\alpha_s f_{\psi})^4$,
where $\alpha_s$ is the strong coupling constant and $f_{\psi}$ is the decay
constant of $\Jp$, in Ref.~\cite{Ecclestone:1982yt} it was found that most of
the measured cross section is due to $q \bar q \to \Jp \Jp$. However, as
mentioned above, such a high production rate is unexpected at NA3 energies.
Therefore, it is interesting to analyze the production rate instead of the
double $\Jp$ production cross section. 

For our purpose it is sufficient to use the quark--hadron duality principle.
Following Ref.~\cite{Amundson:1995em}, the cross section of quarkonium was
obtained by calculating the production of a $Q \bar Q$ in the small invariant
mass interval between $2m_Q$ and the threshold to produce open heavy-quark
hadrons, $2m_H$.\footnote{For an estimate like this, generalizations of the
method given in Refs.~\cite{Gavai:1994in,Nelson:2012bc,Vogt:2015uba,Ma:2016exq}
are not needed.} The $Q \bar Q$ pair has $3\times\bar{3}=(1+8)$ color
components, consisting of a color-singlet and a color-octet. Therefore, the
probability that a color singlet is formed and produces a quarkonium state is
given by $1/(1+8)$, and the model predicts
\begin{eqnarray}
\sigma(Q\bar Q) = \int_{2m_Q}^{2m_H} dM_{Q \bar Q}
  \frac{d\sigma_{Q \bar Q}}{dM_{Q \bar Q}} \, ,
\end{eqnarray}
where $\sigma_{Q \bar Q}$ is the production cross section of the heavy quark
pairs and $\sigma(Q\bar Q)$ is a sum of production cross sections of all
quarkonium states in the duality interval. For example, in case of charmonium
states one has $\sigma(Q\bar Q)=\sigma(\Jp)+\sigma(\psi(2S))+\ldots$\,.
According to a simple statistical counting, the fraction of the total
color-singlet cross section into a quarkonium state is given by
\begin{equation}\label{XQQ}
\sigma(X) = \frac{1}{9} \cdot \rho_X \cdot \sigma(Q\bar Q)
\end{equation}
($X=\Jp,\psi(2S),\ldots$) with
\begin{equation}
\rho_X = \frac{2 J_X + 1}{\sum_i (2J_i + 1)}\, ,
\end{equation}
where $J_X$ is the spin of the quarkonium state $X$ and the sum runs over all
quarkonium states. In case of the $\Jp$ meson the calculation gives
\begin{equation}\label{rhoJpsi}
\rho_{\Jp} \approx 0.2.
\end{equation}

These formulas can be easy generalized to the calculation of the production 
of double quarkonium,
\begin{equation}
\sigma(\Jp\Jp)=\frac{1}{9}\frac{1}{9}\rho_{\Jp}\rho_{\Jp}
  \cdot\sigma(c \bar c + c \bar c).
\end{equation}
Using the NA3 rate it is easy to estimate that 
\begin{equation}
\frac{\sigma (c \bar c + c \bar c)}{\sigma (c \bar c )} > 10^{-2}.
\end{equation}
Even making the unrealistic assumption that all $c \bar c$ pairs in
$\sigma(c\bar c+c\bar c)$ are lying in the duality interval, the production
rate seems to be absolutely untrusted. In essence, assuming SPS to be the only
or even the leading mechanism for double $\Jp$ production at NA3 energies
should rise skepticism.

\subsection{$\sigma(\Jp\Jp)$ production via the Intrinsic Charm mechanism}
The production of double $\Jp$ based on the intrinsic charm approach is
discussed in detail in Refs.~\cite{Vogt:1995tf,Vogt:1995dn,Koshkarev:2016ket}. 
Perturbative QCD and intrinsic charm contributions populate different regions
where the main statistic is expected (cf.\ the discussion in
Ref.~\cite{Koshkarev:2017txl}). Based on this fact, in Ref.~\cite{Vogt:1995tf}
it was assumed that all the NA3 data came from the intrinsic charm mechanism.
However, as we have shown above this kinematic region is limited due to the
detector acceptance. Therefore such interpretation is too ambitious.

Concluding this chapter it is important to note that none of the mechanisms
discussed, neither pQCD nor the intrinsic charm mechanism, can be interpreted
as the sole production mechanism. In addition, it is not possible to determine
the relative contribution, as both calculations, the pQCD
calculation~\cite{Ecclestone:1982yt} as well as the calculation based on the
intrinsic charm mechanism~\cite{Vogt:1995tf}, are normalized to the NA3 data,
assuming either of these to be the sole production mechanism. 

\section{Double $\Jp$ production\\[-10pt] at the COMPASS experiment} 

\subsection{The COMPASS detector: short description}

COMPASS, a fixed target experiment at CERN, uses the high intensity $\pi^-$
beam of $190\GeV$ at the Super Proton Synchrotron at CERN for Drell--Yan (DY)
measurements to produce charmonium, possible exotic states and dimuons in the
set of polarized targets~\cite{Abbon:2014aex}. The experiment had several DY
runs in 2014, 2015 and 2018. 

The COMPASS DY configuration setup
is quite similar to the NA3 setup. It uses two cylindrical cells (of 55 cm
length and 4 cm in diameter each) of ammonia as a target and a hadron absorber
to reduce the particle flux through the setup. The absorber, made of alumina
and stainless steel with the central tungsten plug, is placed downstream of
the target. The outgoing charged particles are detected by two spectrometers
(Large Angle Spectrometer and Small Angle Spectrometer). At each spectrometer,
the muon identification was accomplished by a system of muon filters. To be
detected, at least two muon candidates from the target region should hit the
trigger hodoscopes of the first spectrometer ($25<\theta_\mu<160\mrad$), or one 
should hit the trigger hodoscopes of the first and the other the trigger
hodoscopes of the second spectrometer ($8<\theta_\mu<45\mrad$). A muon passed
through the peripheral part of the absorber and the material of one of two
muon filters (stainless steel or concrete) loses an energy of about $10\GeV$,
defining the lower limit for its reconstruction.
\begin{center}\begin{figure}
\centerline{\includegraphics[scale=0.55]{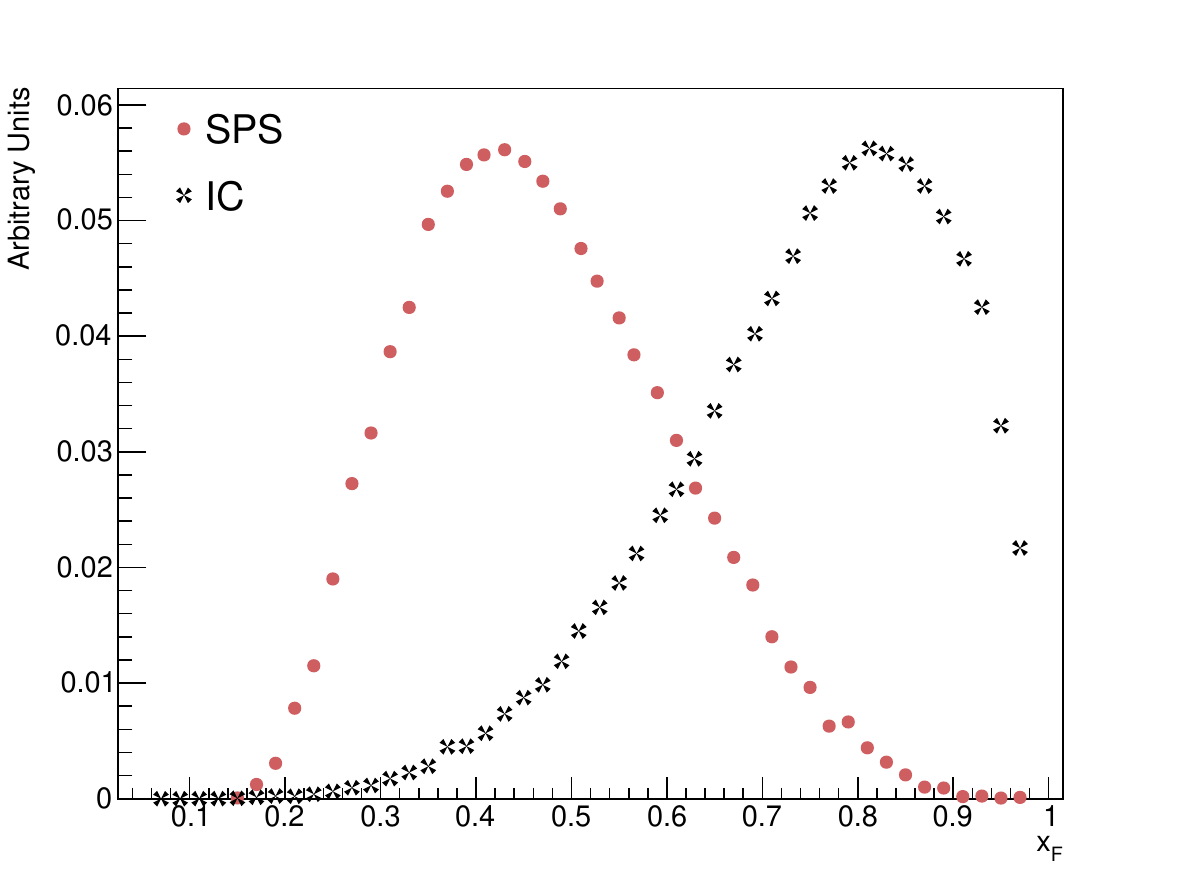}}
\caption{\label{fig:icCompass}Prediction for the $x_F$ distributions for SPS 
and IC production mechanisms. The SPS distribution is obtained by using
HELAC-Onia~\cite{Shao:2012iz, Shao:2015vga}, and the IC distribution is
obtained following Refs.~\cite{Vogt:1995tf,Koshkarev:2016ket}. All
distributions are normalized to unity.}
\end{figure}\end{center}

\subsection{Double $\Jp$ production at the COMPASS}

As the COMPASS has a similar detector setup as the one at NA3, we can estimate
that double $\Jp$ events detected by COMPASS should have $x_F>0.4$ as
threshold. Therefore, COMPASS can give a significant contribution to the
understanding of the double $\Jp$ production mechanisms. In 2015 the COMPASS
collaboration collected about one million dimuon events in the ${\rm NH}_3$
target~\cite{Aghasyan:2017jop}, and a factor of at least $1.5$ more events
are expected in the 2018 run~\cite{Riedl:2018}. Comparing the $\Jp$ statistics
collected by NA3~\cite{Badier:1983dg} and the acceptance of the COMPASS
detector one can estimate up to 100 double $\Jp$ events for the COMPASS
experiment.

Taking into account that perturbative QCD and intrinsic charm contributions
have principally different slopes and different regions where the main
statistics is expected, we propose to use $x_F$ for the search for signals of
the intrinsic charm mechanism and the determination of the relative
contribution (cf.\ discussion in Chapter~\ref{chapter:sps_ic}). The expected
$x_F$ distribution for the COMPASS kinematics and for the different production
mechanisms are shown in Fig.~\ref{fig:icCompass}.

As we already mentioned above, it is not possible to predict the relative 
contribution of the mechanism. However, it is interesting to estimate 
upper limits for both SPS and IC mechanism. Note that Double Parton
Scattering (DPS) contributions are suppressed compared to both SPS and
IC~\cite{Koshkarev:2019crs}.

\subsubsection{Single Parton Scattering}

\begin{center}\begin{figure}
\centerline{\includegraphics[scale=0.55]{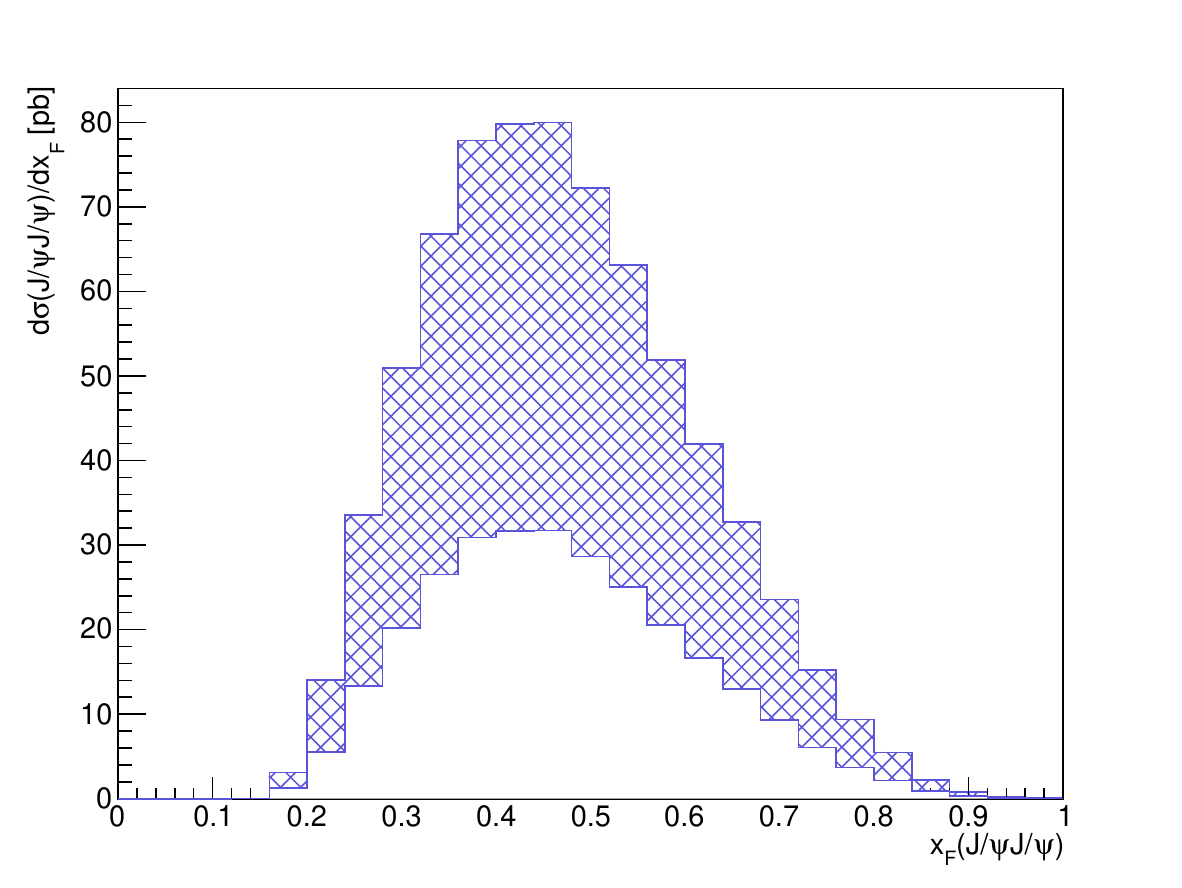}}
\caption{\label{fig:spsCompass}Prediction for the $x_F$ distributions for 
SPS. The shape of the SPS distribution is obtained by using
HELAC-Onia~\cite{Shao:2012iz, Shao:2015vga}. The region $x_F < 0.4$ is
excluded by the COMPASS acceptance.}
\end{figure}\end{center}

Following calculations of the double $\Jp$ production cross section in SPS
from Ref.~\cite{Humpert:1983qt} we can find a ratio between the double $\Jp$
production cross sections with a $\pi^-$ beam at NA3 and COMPASS energies:
\begin{equation}
\sigma_{\Jp\Jp}(150\GeV /c) : \sigma_{\Jp\Jp}(190\GeV /c) : \sigma_{\Jp\Jp}(280\GeV /c) 
\approx 1: 2.06 : 3.34 \, .
\end{equation}
Using the mean values for the double $\Jp$ production cross sections measured
by NA3 of $18 \pm 8\pb$ and $30 \pm 10\pb$ per nucleon at $150$ and
$280\GeV/c$ as  reference points, we find
$\sigma(\Jp\Jp) \approx (12-29) \pb$ per nucleon at $190\GeV/c$ 
(cf.\ Fig.~\ref{fig:spsCompass}).

\subsubsection{Intrinsic Charm}

\begin{center}\begin{figure}
\centerline{\includegraphics[scale=0.55]{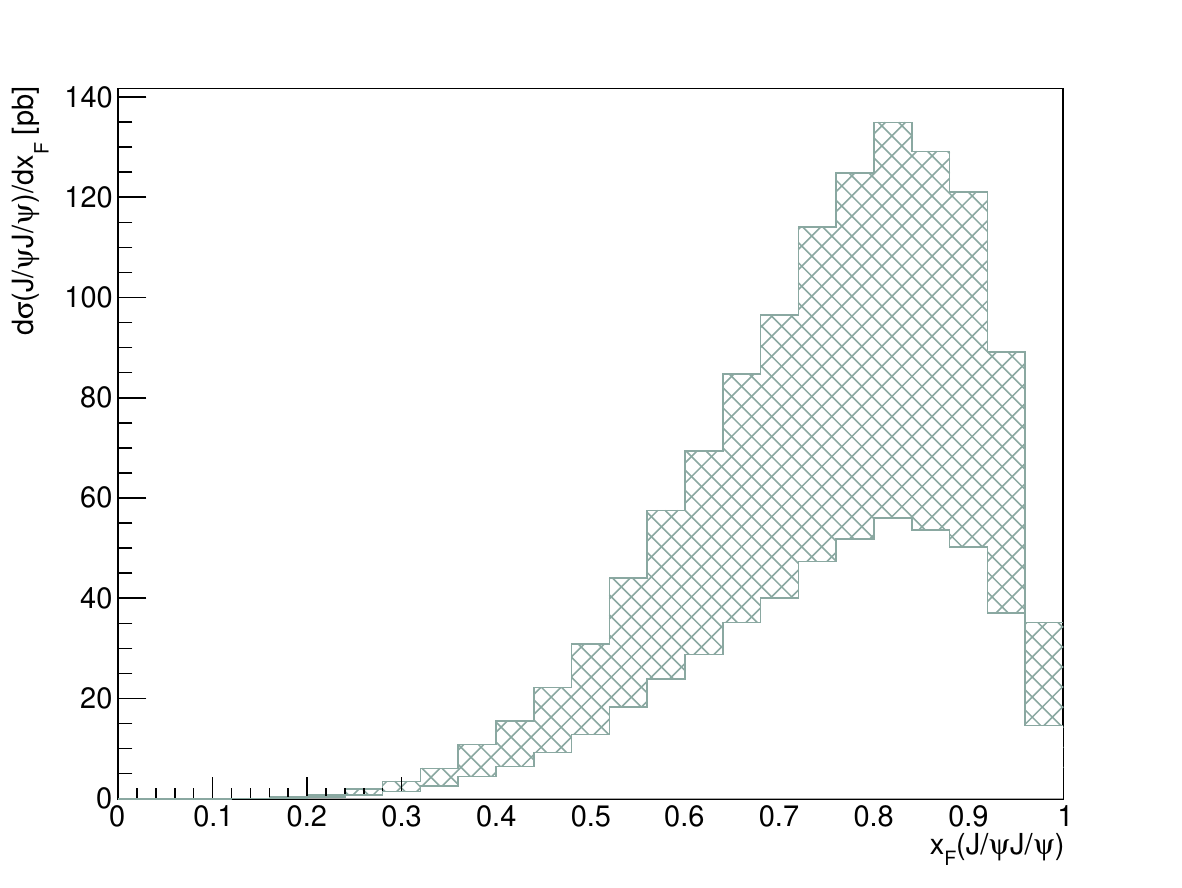}}
\caption{\label{fig:vogt}Prediction for the $x_F$ distributions for the
intrinsic charm mechanism. The region $x_F < 0.4$ is excluded by the COMPASS
acceptance.}
\end{figure}\end{center}

Following Ref.~\cite{Vogt:1995tf}, we cast the double $\Jp$ production 
cross section into the form
\begin{equation}
\sigma_{\Jp\Jp} = f^2_{\psi/\pi} \frac{P_{icc}}{P_{ic}} \sigma_{ic},
\end{equation}
where $f_{\psi/\pi} \approx 0.03$ is the fraction of $c \bar c$ quark pairs 
producing $\Jp$, $P_{ic}$ and $P_{icc}$ are probabilities to produce 
intrinsic $c \bar c$ and $c \bar c c \bar c$ Fock states, respectively, and 
$\sigma_{ic} \approx 0.5 \mub$ is the intrinsic charm cross section for a
$\pi^-$ beam momentum of $200 \GeV/c$. Assuming $P_{icc}$ to be independent of
the projectile, the value $P_{icc} = 4.4\% P_{ic}$ was found in
Ref.~\cite{Vogt:1995tf}, and assuming $\sigma(\Jp\Jp)/\sigma(\Jp)$ to be
independent of the projectile, $P_{icc} = 10.6\% P_{ic}$ was found in the same
Ref.~\cite{Vogt:1995tf}. It is easy to estimate the double $\Jp$ production
cross section at the COMPASS energy to be $(19.8-47.7) \pb$ per nucleon
(cf.\ Fig.~\ref{fig:vogt}).

\subsection{Combination of Single Parton Scattering\\
  and Intrinsic Charm contributions\label{chapter:sps_ic}}

\begin{figure}
\includegraphics[scale=0.43]{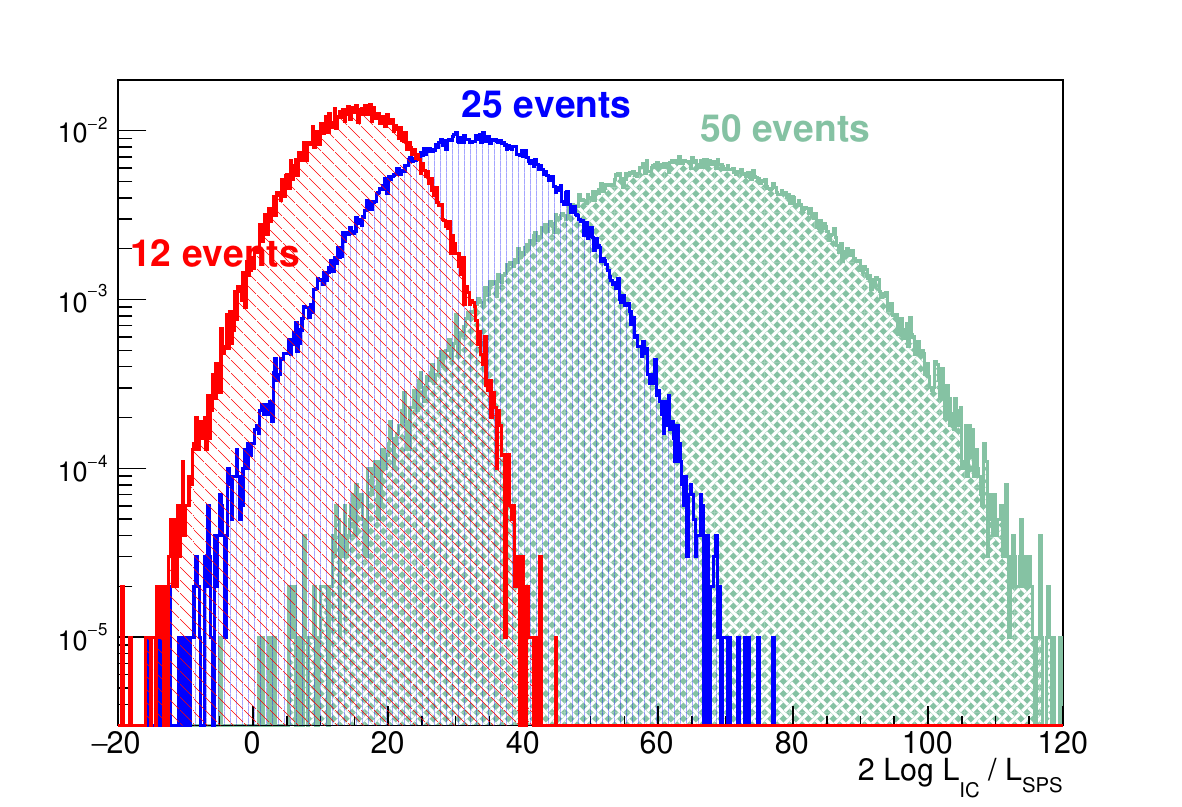}\kern-16pt
\includegraphics[scale=0.43]{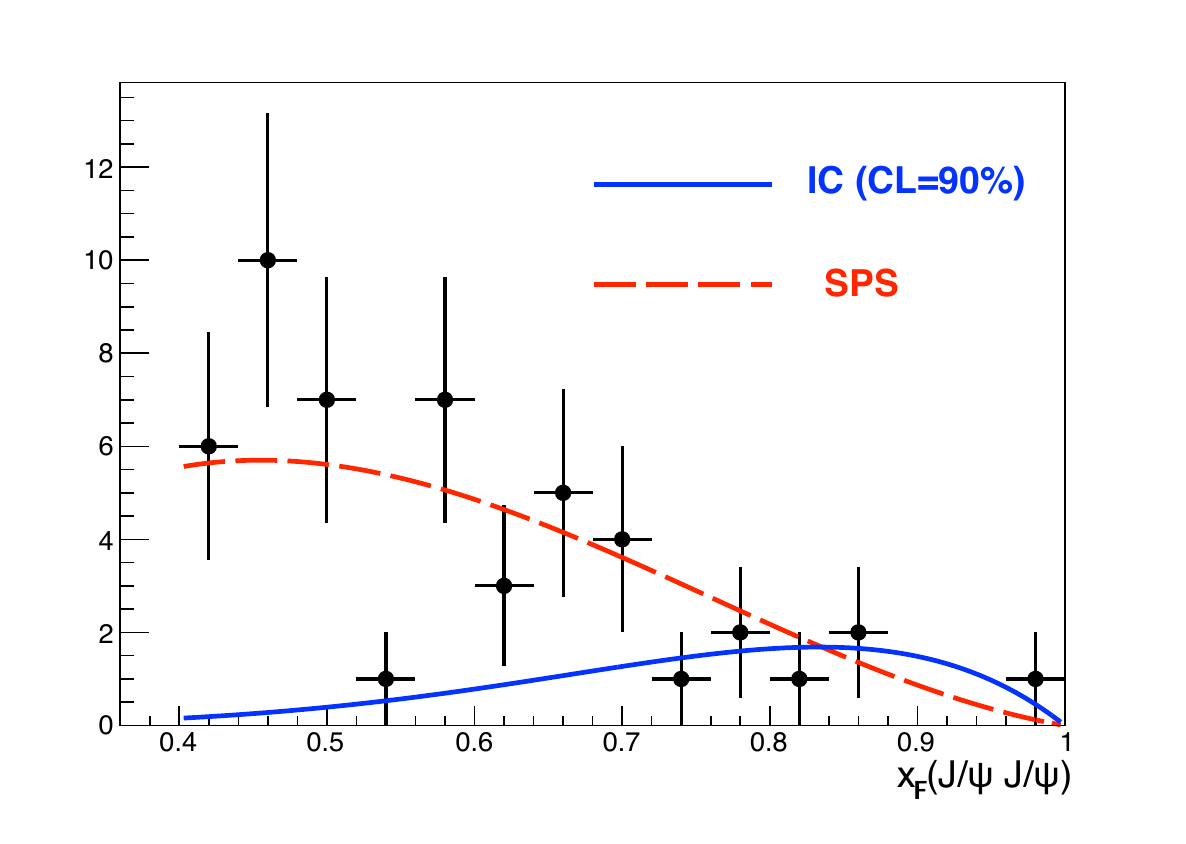}\kern-18pt
\caption{\label{fig:upperlimit} Double likelihood ratio distribution as a
measure of the COMPASS separation power for the double $\Jp$ production via IC
and SPS mechanisms for the statistics of 12, 25 and 50 events (left panel) and
a toy Monte Carlo result for double $\Jp$ production via SPS together with an
upper limit for the IC mechanism contribution for the statistics of 50 events
(right panel).}
\end{figure}

As all researchers before us, up to this point we more or less quietly assumed
the only possible production mechanism to be either SPS or DPS. However, it is
interesting to investigate the case where both mechanisms can contribute to
the production of double $\Jp$ and to investigate the ability of the COMPASS 
experiments to separate them.

The possibility for COMPASS to identify the IC production mechanism based on
the measured $x_F(\Jp\Jp)$ distribution is investigated using a toy Monte
Carlo simulation where events are generated according to the SPS contribution
shown in Fig.~\ref{fig:icCompass}. The flat acceptance in the range
$0.4<x_F(\Jp\Jp)<1$ and the absence of any other contributions is assumed. The
double likelihood ratio distribution $R=2\log(L_{\rm IC}/L_{\rm SPS})$ based
on $10^5$ runs is shown in Fig.~\ref{fig:upperlimit} (left) for the COMPASS
statistics of 12, 25 and 50 events. The likelihoods $L_{\rm IC}$ and
$L_{\rm SPS}$ correspond to double $\Jp$ production via pure IC and pure SPS
mechanisms, respectively. In case of 50 double $\Jp$ events an upper limit
(CL=90\%) for the relative IC mechanism contribution to the production cross
section in the range $x_F(\Jp\Jp)>0.4$ could be established on the level of
29\%. The obtained numbers should be treated just as an illustration as a
non-flat acceptance and/or the presence of any additional significant
contributions (e.g.\ background) could change these numbers significantly in
both directions. The result of a single Monte Carlo run with the fitted SPS
contribution and the curve corresponding to the obtained upper limit
contribution of IC are shown as an example in Fig.~\ref{fig:upperlimit}
(right). 

\section{Conclusions}

The analysis of this paper clearly shows that the NA3 data is puzzling and
does not allow for a simple interpretation. We found that the calculation of
the double $\Jp$ production cross section is not very useful for the
identification of the production mechanism. Instead, we have pointed out that
the kinematic distributions provide opportunities for the COMPASS experiment
using the $\pi^-$ beam of the Super Proton Synchrotron (CERN) at $190\GeV/c$
to measure the effect of the intrinsic charm mechanism or to identify the
production mechanism. We also estimated the double $\Jp$ production cross
section in assumption that all NA3 events came either from the SPS or from the
IC mechanism. In addition, we give an estimate for the double $\Jp$ events
to be expected at COMPASS.

\subsection*{Acknowledgements}
We would like to thank H.S.~Shao for updating the HELAC-Onia generator. Also
we would like to thank D.~Bandurin for useful discussions. This research was
supported by the Estonian Research Council under Grants No.~TK133 and~PRG356.

\end{document}